\documentclass[aps,prl,reprint,superscriptaddress,nofootinbib,longbibliography,floatfix]{revtex4-2}
\usepackage[T1]{fontenc}
\usepackage[utf8]{inputenc}
\usepackage{lmodern,microtype}
\usepackage{amsmath,amssymb,bm,mathtools}
\usepackage{graphicx,booktabs,xcolor,hyperref}
\hypersetup{colorlinks=true,linkcolor=blue!45!black,citecolor=blue!45!black,
urlcolor=blue!55!black}
\newcommand{\Id}{\mathrm I}
\newcommand{\E}{\mathbb E}
\newcommand{\Tr}{\operatorname{Tr}}
\newcommand{\dd}{\mathrm d}
\newcommand{\ii}{\mathrm i}
\newcommand{\sech}{\operatorname{sech}}
\newcommand{\diag}{\operatorname{diag}}
\newcommand{\norm}[1]{\left\lVert #1\right\rVert}

\graphicspath{{figures/}}
\begin{document}
\title{Full Lyapunov Spectra of Random Neural Networks,\\Part I: Finite-Time and Asymptotic Theory}
\author{Haoyu Albert Wang}
\affiliation{Research Institute of Intelligent Complex Systems, Fudan University, Shanghai, China}
\affiliation{Institute of Science and Technology for Brain-Inspired Intelligence, Fudan University, Shanghai, China}
\author{Wei P. Dai}
\affiliation{Research Institute of Intelligent Complex Systems, Fudan University, Shanghai, China}
\begin{abstract}
We develop a dynamical mean-field framework for the full finite-time
and stationary asymptotic Lyapunov spectra of random recurrent neural
networks. At finite observation time, an exact log-determinant
representation of tangent-propagator singular values leads to a
closed-contour matrix Dyson equation for the smoothed spectral tail.
In the stationary setting, a cocycle representation yields an operator
equation for the asymptotic cumulative spectrum. Both formulations
retain complete local gain histories and the repeated action of
quenched connectivity. Direct singular-value calculations and network
QR spectra support the predictions in the regimes examined, with
sampling uncertainty and numerical cutoffs assessed separately.
The stationary equation also recovers the known maximal exponent,
and controlled gain-history replay illustrates how temporal
correlations redistribute instability. This work was carried out in
collaboration with ChatGPT 5.6 Sol.
Toward the end of preparation we became aware of David G. Clark's contemporaneous
work, \emph{Lyapunov spectrum of random neural networks}
[arXiv:2610.12426], which reports overlapping mean-field results for the
full Lyapunov spectrum using a cavity approach.
\end{abstract}
\maketitle
\paragraph{Introduction.}
The full Lyapunov spectrum describes how perturbations grow or contract
across the tangent space of a dynamical system. At finite observation
times, the singular values of the tangent propagator define a spectrum
of amplification rates; their long-time limits determine the asymptotic
Lyapunov spectrum. In recurrent neural networks, these distributions
connect finite-time amplification to the number of unstable directions
and the growth of tangent volumes. For the random rate network of
Sompolinsky, Crisanti, and Sommers \cite{SCS1988}, classical dynamical
mean-field theory (DMFT) determines stationary activity and the maximal
exponent \cite{CrisantiSompolinsky2018}. Numerical studies have also
established an extensive full spectrum and its relation to entropy
production and the Kaplan--Yorke dimension \cite{Engelken2023}.
The theoretical task is to describe the entire spectrum at finite
times and in the stationary asymptotic regime.

The central difficulty lies in the product structure of the tangent
propagator. Local gains evolve along a nonlinear trajectory, while the
same quenched connectivity acts at every time. The resulting temporal
dependence is essential: a one-time Jacobian distribution does not specify
the product law. Our formulation retains both ingredients. DMFT supplies
a probability law for entire gain histories, and a Hermitized matrix
Dyson equation (MDE) averages the connectivity while preserving its reuse
throughout the evolution.

We formulate these two spectral problems within a common mean-field
framework. At finite time, an exact log-determinant identity leads to a
closed-contour MDE for the smoothed singular-growth-rate spectrum.
For stationary dynamics, a cocycle representation yields an operator
MDE for the asymptotic spectral tail. Direct propagator calculations
test the finite-time theory, while network QR spectra test the
stationary predictions with sampling and numerical uncertainties
assessed separately. The comparisons distinguish independently
prescribed gain histories from autonomous network trajectories, whose
gains are generated by the connectivity that also governs tangent
growth. Recovery of the known maximal exponent provides a consistency
check of the stationary theory; controlled replay then illustrates how
temporal memory shapes the distribution of instability.

Toward the end of preparation we became aware of Clark's recent work
\cite{Clark2026} on the Lyapunov spectrum of random neural networks.
Clark relates the cumulative exponent distribution to the response of
shifted tangent dynamics selected by a minimum-norm condition. A
regularized forward--backward formulation and a cavity calculation then
yield a self-consistent single-site description at large network size.
Numerical solutions are compared with network simulations, and the
resulting spectrum is used to study the Kaplan--Yorke dimension and the
entropy rate, characterizing the extensivity of chaos.

\paragraph{Model and spectral observables.}
We set the single-neuron relaxation time to one and study
\begin{align}
 \dot x_i&=-x_i+\sum_j W_{ij}\tanh x_j, \label{eq:model}\\
 \dot u&=[-\Id+WD(t)]u,\qquad D_{ii}(t)=\sech^2x_i(t),
 \label{eq:tangent}
\end{align}
where $W_{ij}=g\xi_{ij}/\sqrt N$, $W_{ii}=0$, and the off-diagonal
$\xi_{ij}$ are independent standard real Gaussians. On the stationary
chaotic DMFT branch, $x(t)$ is a zero-mean Gaussian process satisfying
\begin{equation}
 (1-\partial_\tau^2)C_x(\tau)
 =g^2\E[\tanh x(0)\tanh x(\tau)]. \label{eq:activity}
\end{equation}
It induces a generally non-Gaussian gain process $d(t)=\sech^2x(t)$.
The spectral calculation uses its path law, not a fit to Lyapunov data.

Let $M_T$ be the tangent propagator, with singular values
$s_i(M_T)=\exp(T\lambda_i^{(T)})$. Define the spectral free energy
\begin{equation}
 F_{N,T}(\mu)=\frac{1}{2NT}\log\det
 [\Id+e^{-2\mu T}M_T^\dagger M_T]. \label{eq:finite-free}
\end{equation}
Its derivative is the exact soft spectral tail
\begin{equation}
 \nu_{N,T}(\mu)=-\partial_\mu F_{N,T}(\mu)
 =\frac1N\sum_i\frac{1}{1+e^{2T(\mu-\lambda_i^{(T)})}}.
 \label{eq:finite-tail}
\end{equation}
The smoothing width is $O(T^{-1})$. The determinant also generates
tangent-volume growth at every rank:
\begin{equation}
 \det(\Id+qM_T^\dagger M_T)
 =\sum_{k=0}^{N}q^k\norm{\wedge^kM_T}_{\mathrm F}^2.
 \label{eq:exterior}
\end{equation}
These identities hold at finite $N$ for any propagator.

At fixed $N$, the multiplicative ergodic theorem \cite{Oseledets1968}
gives the long-time limit. When the subsequent empirical-measure limit
has the required first-moment integrability,
\begin{align}
 F(\mu)&=\int(\lambda-\mu)_+\rho(\dd\lambda),
 &\nu(\mu)&=-F'(\mu), \label{eq:limit-free}\\
 p_+&=\nu(0),&h_+&=F(0). \label{eq:observables}
\end{align}
Here $p_+$ is the positive-exponent fraction and $h_+$ is the positive
Lyapunov sum per neuron. At atoms, $\nu$ denotes the right-continuous
hard tail; the finite-time smoothing need not select that convention
exactly at an atom. The measure $\rho=F''$ is defined distributionally.
When the Kaplan--Yorke crossing is interior and continuous,
\begin{equation}
 F(\mu_*)+\mu_*\nu(\mu_*)=0,\qquad
 d_{\rm KY}=\nu(\mu_*). \label{eq:ky}
\end{equation}
The quantity $h_+$ equals the Kolmogorov--Sinai entropy per neuron under
the hypotheses of Pesin's identity \cite{Pesin1977}; $d_{\rm KY}$ is a
spectral estimate of attractor dimension \cite{KaplanYorke1979}.

\paragraph{Finite-time path equation.}
Use the discretization
\begin{equation}
 \begin{gathered}
 B_t=c\Id+bWD_t,\quad c=e^{-\delta},\quad b=1-e^{-\delta},\\
 M_T=B_{m-1}\cdots B_0,\quad T=m\delta.
 \end{gathered}
 \label{eq:step}
\end{equation}
It treats the leak exactly and remains linear in $W$. The forward factors
$B_0,\ldots,B_{m-1}$ followed by their adjoints in reverse order form a
$2m$-step contour. A block-cyclic linearization $\mathbb L_\mu$ obeys
\begin{equation}
 \det\mathbb L_\mu
 =\det(\Id+e^{-2\mu T}M_T^\dagger M_T). \label{eq:linearization}
\end{equation}
Hermitization \cite{FeinbergZee1997} produces site blocks of size $4m$.
Their deterministic part is $h_0$, and the path-dependent coefficient of
the Gaussian coupling is $A_d$. Appendix~\ref{app:finite} gives all
indices, including the boundary sign.

For gain histories sampled independently across sites and independently
of $W$, the two disorder contractions yield
\begin{align}
 \mathcal M_d^{-1}(z)&=h_0-z\Id-g^2(X+A_d^\dagger\overline{\mathcal M}A_d),
 \label{eq:finite-mde}\\
 \overline{\mathcal M}&=\E_d\mathcal M_d,\qquad
 X=\E_d(A_d\mathcal M_d A_d^\dagger). \label{eq:finite-orders}
\end{align}
The physical resolvent branch has $\Im\mathcal M_d\succeq0$ for
$z=\ii\eta$, $\eta>0$. The required tail is a corner readout,
\begin{equation}
 \nu_T(\mu)=1-\lim_{\eta\downarrow0}
 \Re\E_d[\mathcal M_d(\ii\eta)]_{2m,0}. \label{eq:finite-readout}
\end{equation}
The covariance structure keeps the correlations created by reuse of the
same quenched matrix. Averaging the gains before the inverse would give
a different equation.

\begin{figure*}[t]
\centering\includegraphics[width=.96\textwidth]{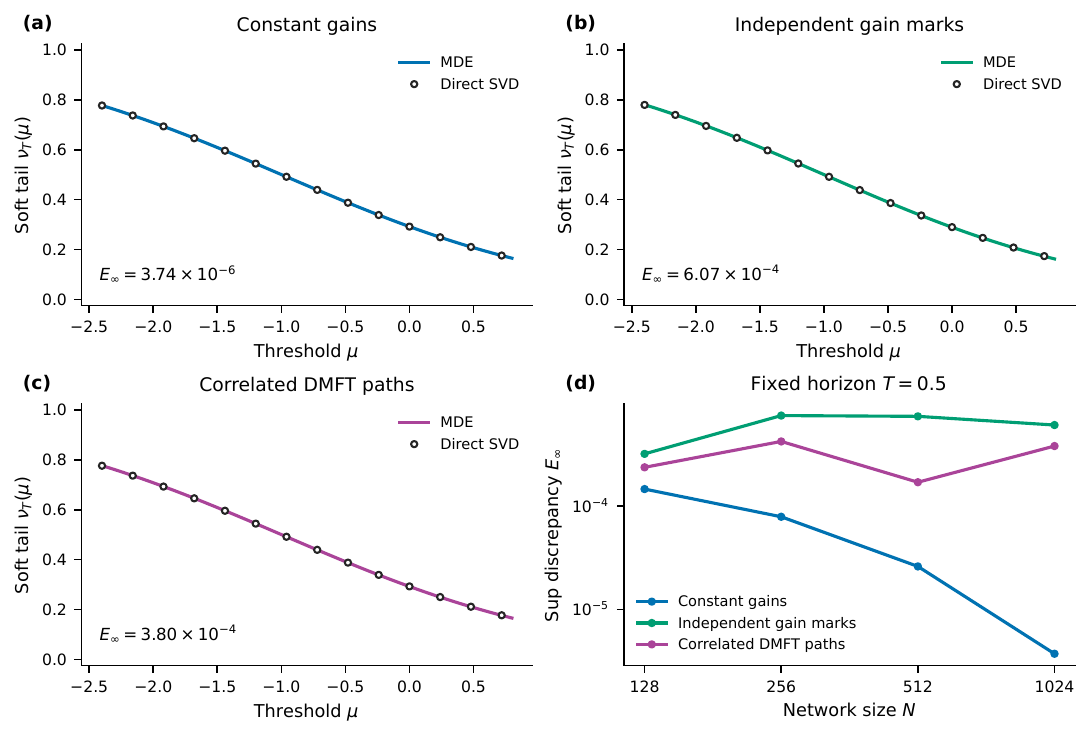}
\caption{Finite-time contour MDE versus direct singular values.
(a)--(c) Constant, independent, and correlated gains at $g=1.5$,
$\delta=0.05$, $T=0.5$, $N=1024$. The curves use the same logistic
observable, Eq.~\eqref{eq:finite-tail}; eight direct realizations are
averaged. (d) Size controls. The weights are quenched in every direct
product. These panels test independent prescribed paths.}
\label{fig:finite}
\end{figure*}

Direct products recover the predicted soft tails for constant,
temporally independent, and DMFT-correlated gains
(Fig.~\ref{fig:finite}). These protocols test the treatment of temporal
structure while retaining the same quenched matrix throughout each
product. Longer replay tests span multiple horizons, system sizes,
integration steps, and replay speeds. All 45 predictions lie within their respective
pointwise 95\% Student intervals for the direct mean. This is sampling
compatibility, not simultaneous 95\% coverage or uniform percent accuracy:
the largest relative error in a thin tail is 23.15\%.

\paragraph{Finite-time validation across coupling strengths.}
We next compare with gain histories generated by the autonomous network
itself. The comparison spans $g=1.2,1.5,3$ and
$T=0.5,1,2,4,8,16$, with $\delta=0.05$ and
$N=256,512,1024$; the $g=1.5$ series also includes $N=2048$.
For each matrix, one propagator follows the network trajectory and a
second follows independent Gaussian DMFT histories with the same
connectivity. Both use Eq.~\eqref{eq:step}, and both are measured by
direct SVD. Thus the comparison concerns the finite-time soft tail in
Eq.~\eqref{eq:finite-tail}, not a prefix of QR exponents.

Let $\bar\nu_T^{\rm net}$ and $\bar\nu_T^{\rm path}$ denote the
respective connectivity-ensemble means. On the common threshold grid
$\mathcal G$, we separate
\begin{align}
 E_A&=\max_{\mu\in\mathcal G}|\nu_T^{\rm MDE}-\bar\nu_T^{\rm path}|,
 \nonumber\\
 E_B&=\max_{\mu\in\mathcal G}|\bar\nu_T^{\rm path}-\bar\nu_T^{\rm net}|,
 \nonumber\\
 E_{\rm net}&=\max_{\mu\in\mathcal G}|\nu_T^{\rm MDE}-\bar\nu_T^{\rm net}|.
 \label{eq:finite-network-errors}
\end{align}
The first quantity probes the marked-path equation; the second probes
replacement of autonomous paths by independent DMFT paths. The third
tests the combined network prediction, and obeys
$E_{\rm net}\leq E_A+E_B$.

\begin{figure*}[t]
\centering\includegraphics[width=.98\textwidth]{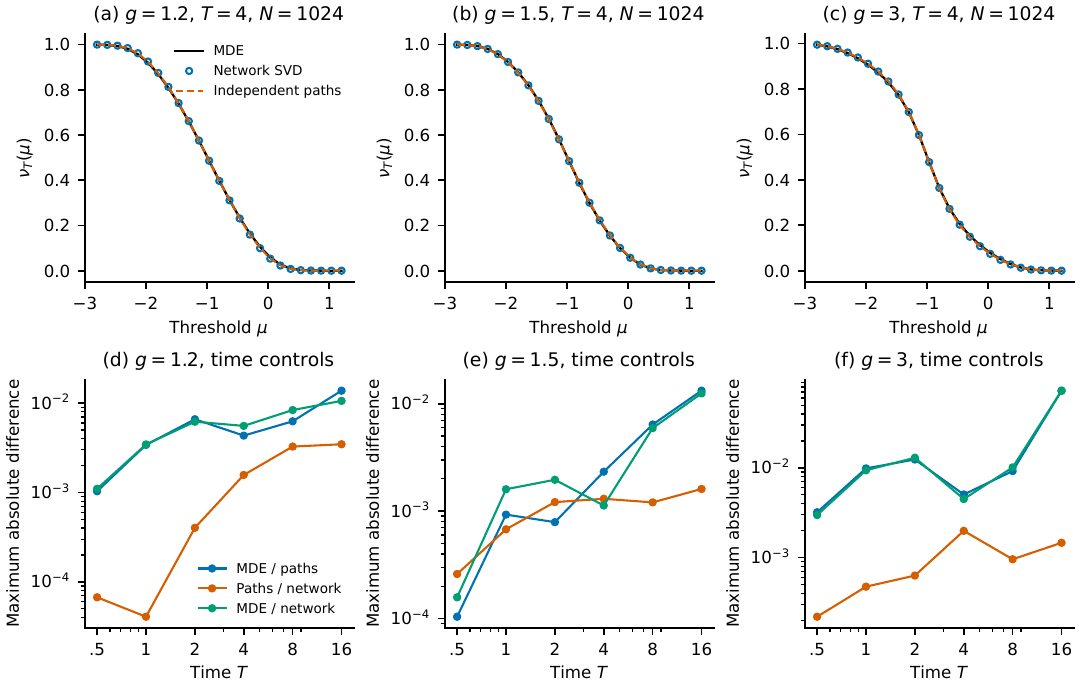}
\caption{Finite-time autonomous-network validation at three couplings.
(a)--(c) $T=4$, $N=1024$: MDE predictions, direct network SVD means,
and independent-path SVD means for the same soft observable.
(d)--(f) The three errors in Eq.~\eqref{eq:finite-network-errors} at all
six observation times, using the largest available path population at
each time. Path counts vary with time, as specified in
Appendix~\ref{app:numerics}; the rise of an error therefore does not
isolate a physical time effect. No sampling intervals are implied by the
connecting lines.}
\label{fig:finite-network}
\end{figure*}

The predicted and measured soft-tail curves agree closely at the common
comparison horizon (Fig.~\ref{fig:finite-network}). At $T=4,N=1024$,
the network discrepancy remains below $0.0056$ for every coupling.
The time controls show how the errors evolve with observation horizon
and path resolution; Table~\ref{tab:finite-network} gives their
decomposition and sample counts. The additional $g=1.5,N=2048$ point
uses only one matrix, so it cannot establish monotone size convergence
or a sampling interval. These comparisons support the finite-time
construction while leaving time-uniform accuracy as a separate question.

\begin{table*}[t]
\caption{Finite-time network comparisons on the 25 threshold nodes. $n_W$ counts independent matrices; $R$ counts MDE paths per population. $E_A$, $E_B$, and $E_{\rm net}$ are defined in the text. The $N=2048$, $T=4$ entry has only one matrix and provides no sampling interval.}
\label{tab:finite-network}
\begin{ruledtabular}\begin{tabular}{rrrrrrrr}
$g$ & $T$ & $N$ & $n_W$ & $R$ & $E_A$ & $E_B$ & $E_{\rm net}$\\
1.2&4&1024&2&16&0.004324&0.001570&0.005553\\
1.5&4&1024&3&64&0.002332&0.001304&0.001129\\
1.5&4&2048&1&64&0.005872&0.001006&0.006502\\
3&4&1024&2&16&0.005045&0.001985&0.004493\\
\end{tabular}\end{ruledtabular}
\end{table*}

\paragraph{Stationary spectrum.}
For a stationary ergodic gain environment, let $U$ implement the time
shift and $Y=BU$ represent the tangent cocycle. With normalized trace
$\tau$, Deninger's determinant formula \cite{Deninger2011} gives
\begin{equation}
 F_N(\mu)=\delta^{-1}\log\Delta_\tau(e^{\delta\mu}\Id-Y_N)-\mu.
 \label{eq:brown-free}
\end{equation}
The determinant is the Fuglede--Kadison determinant
\cite{FugledeKadison1952}. Under the cocycle integrability assumptions,
its radial Brown measure has logarithmic radii $\delta\lambda_i$.
Equation~\eqref{eq:brown-free} is an exact representation at fixed $N$;
it does not by itself establish a large-$N$ neural closure.

In a time representation, set $(Sf)_t=f_{t-1}$,
$r=e^{\delta\mu}$, $\ell_r=r\Id-cS$, and $K_d=-bD_dS$.
The stationary version of the same covariance contraction is
\begin{equation}
 \mathcal M_d(r,z)=
 \begin{pmatrix}
 -z\Id-\Gamma&\ell_r\\
 \ell_r^\dagger&-z\Id-g^2K_d^\dagger P K_d
 \end{pmatrix}^{-1}, \label{eq:stationary-mde}
\end{equation}
with
\begin{equation}
 P=\E_dG_{++,d},\qquad
 \Gamma=g^2\E_d(K_dG_{--,d}K_d^\dagger), \label{eq:stationary-orders}
\end{equation}
where $G_{\alpha\beta,d}$ are the blocks of $\mathcal M_d$.
The tail is
\begin{equation}
 \nu(\mu)=1-r\lim_{\eta\downarrow0}
 \Re\tau_{\rm erg}\E_dG_{-+,d}(r,\ii\eta). \label{eq:stationary-readout}
\end{equation}
Stationarity makes $P$ and $\Gamma$ convolution operators, represented
by two Fourier functions. The individual path-conditioned inverse
continues to depend on the entire gain history. The numerical period
$L\delta$ truncates a stationary operator; it is not a QR observation
time.

We use Eq.~\eqref{eq:stationary-mde} as a numerically tested closure.
Its prescribed-path Gaussian derivation is separate from its use for the
autonomous network. The latter requires a stationary cavity replacement
and control of small singular values in the zero-regulator limit.
Neither a small equation residual nor agreement of a coarse full
spectral curve proves these assumptions uniformly near a spectral edge.

\begin{figure*}[t]
\centering\includegraphics[width=.96\textwidth]{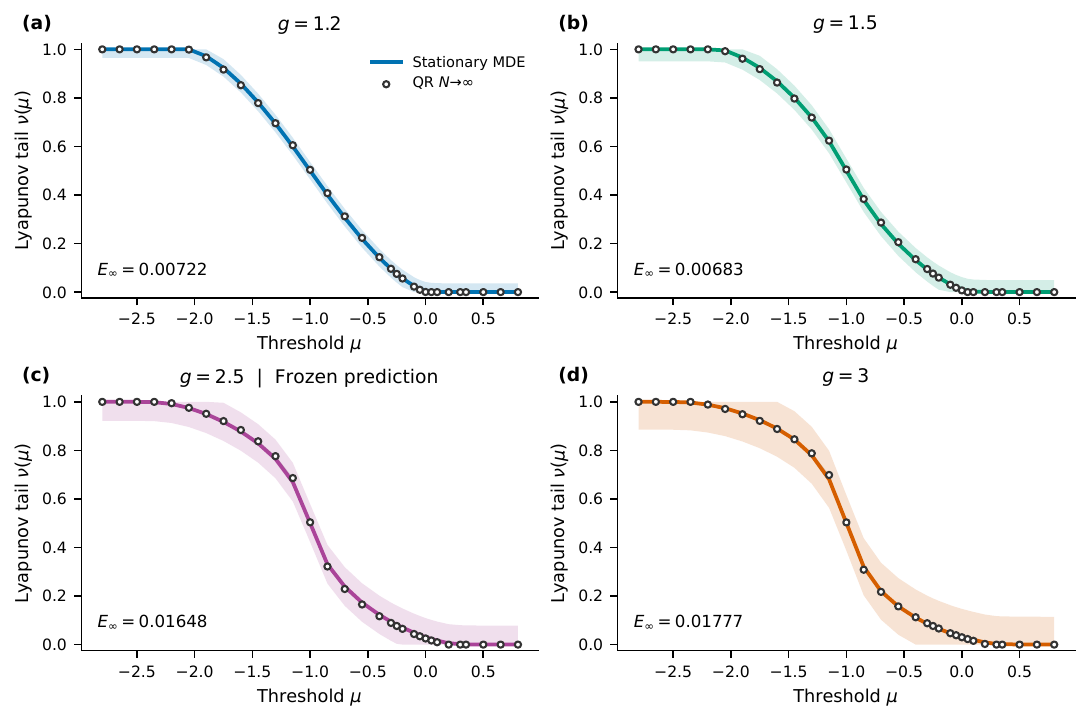}
\caption{Stationary spectral tails and autonomous-network QR references.
Lines are MDE predictions, symbols are finite-size-extrapolated QR
values. Colored bands combine MDE path uncertainty and numerical
systematics; gray bands show pointwise QR sampling intervals.
The $g=2.5$ prediction was recorded before its reference simulations.
Compatibility of these full curves does not establish relative precision
in the sparse extreme tail.}
\label{fig:stationary}
\end{figure*}

\paragraph{Full-spectrum validation.}
Figure~\ref{fig:stationary} compares stationary predictions with network
QR spectra over a range of coupling strengths. The theory captures the
overall spectral shape, with maximum absolute tail discrepancies below
$0.018$, within the conservative sampling and systematic-error budgets.
The $g=2.5$ prediction and reconstruction rules were frozen before its
reference simulations. Appendix~\ref{app:numerics} summarizes the
uncertainty budgets, numerical cutoff controls, and separate
direct-matrix and paired network/path tests.

Resolving the positive sector requires tighter control of regularizer
leakage. A subsequent test using implicit differentiation brings the
positive-exponent fraction, positive Lyapunov sum, and Kaplan--Yorke
dimension into agreement with QR readouts to within 3\%
(Fig.~\ref{fig:positive}). This comparison uses fixed numerical cutoffs
and was performed after the QR reference was available; its precision
does not constitute a bound on all continuous-time limiting observables.
The numerical settings and limits of this refinement are given in
Appendix~\ref{app:numerics}.

\paragraph{Recovery of the maximal exponent.}
The maximal exponent follows from the exterior stability of the full
stationary MDE. As the zero-regulator exterior branch loses stability,
its linearized equation reduces to a one-dimensional Schr\"odinger
problem with effective potential $1-g^2C_d(\tau)$, where
$C_d(\tau)=\E[d(0)d(\tau)]$. If $E_0$ is the lowest spectral value of
$\mathcal H=-\partial_\tau^2+1-g^2C_d(\tau)$, the outer boundary is
\begin{equation}
 \lambda_e=-1+\sqrt{1-E_0}. \label{eq:edge-value}
\end{equation}
This recovers the established two-copy DMFT result
\cite{SCS1988,CrisantiSompolinsky2018}. At $g=1.2,1.5,2.5$, the
step-extrapolated MDE edges are $0.011919464$, $0.046745834$, and
$0.175779366$, respectively. Their absolute differences from independent
two-copy calculations using the same activity-DMFT input are
$9.51\times10^{-8}$, $4.02\times10^{-7}$, and $2.35\times10^{-6}$.
For this model, the linear
boundary equation uses only the two-time gain correlation, while the
nonlinear bulk equation retains the full gain-path law. This consistency
check does not determine the local spectral density near the edge.
Appendix~\ref{app:edge-reduction} gives the reduction and its interpretation;
Appendix~\ref{app:edge-numerics} gives volume and step controls for
the MDE--two-copy comparison.

\paragraph{Isolating temporal memory.}
Consider the independently prescribed gain process and replay it as
\begin{equation}
 d_a(t)=d(at),\qquad
 \dot u=[-\Id+WD_a(t)]u,\qquad a>0. \label{eq:replay}
\end{equation}
The leak time is held fixed. Stationarity makes the one-time gain law,
and hence the one-time distribution of $-\Id+WD_a(t)$, independent of
$a$. The two-time law changes to $C_{d,a}(\tau)=C_d(a\tau)$.
Moreover, since $W_{ii}=0$, Liouville's identity gives exactly
\begin{equation}
 \frac1N\sum_i\lambda_i=-1 \label{eq:trace}
\end{equation}
for the continuous cocycle at every replay rate. These facts allow a
direct test of memory with instantaneous statistics and total contraction
held fixed.

\begin{figure*}[t]
\centering\includegraphics[width=.94\textwidth]{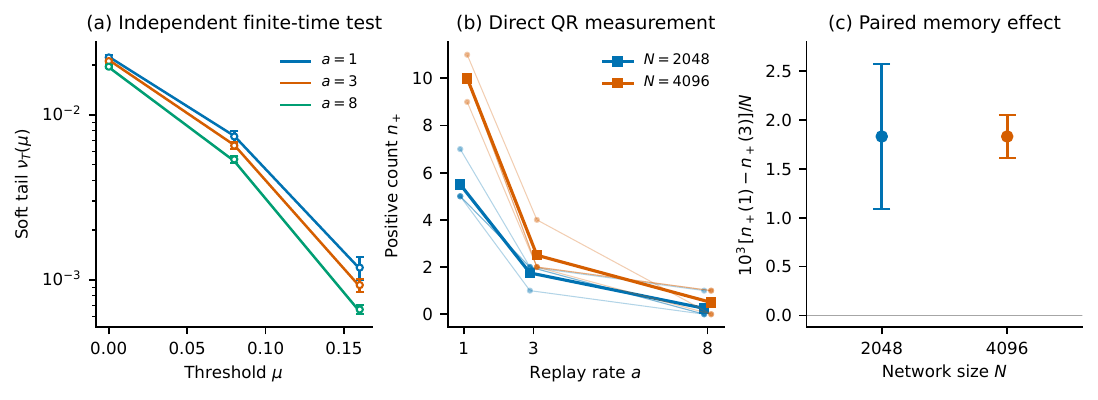}
\caption{Direct tests of prescribed temporal replay at $g=1.3$.
(a) Finite-time soft tails at $N=1024$, $T=16$, $\delta=0.25$;
lines connect MDE values and symbols show 12-realization SVD means
with pointwise 95\% Student intervals. (b) Positive Lyapunov counts
from independent RK4--QR propagation; pale lines connect the same
connectivity across replay rates. Bold lines show means.
(c) The paired decrease from $a=1$ to $a=3$, divided by $N$, with
Student intervals. The latter two panels are measurements, with no MDE
fit. Alignment time is 500 and measurement time is 1500.}
\label{fig:replay}
\end{figure*}

Independent RK4--QR propagation at $g=1.3$, with four matrices per size,
shows a clear reduction in the unstable sector at both studied sizes
(Fig.~\ref{fig:replay}). For example, at $N=4096$, changing $a=1$ to
$a=3$ lowers the mean positive count from $10.0$ to $2.5$. The paired
reductions are resolved at both sizes, and the integration-step control
preserves the positive counts. Appendix~\ref{app:replay} gives the
sampling intervals and numerical checks. Finite-size positive-exponent
outliers remain at the fastest replay rate.

The physical conclusion is the measured memory dependence. Absolute
stationary replay tails are less completely validated: two of the 30
covered comparisons exceed their between-matrix sampling intervals.
Appendix~\ref{app:replay} reports these deviations and the scope of the
sampling intervals. Thus the intervention does not establish a
uniformly accurate stationary closure or a sharp replay transition.
The replay system is a prescribed random linear cocycle, not a newly
constructed autonomous neural attractor; its $h_+$ and $d_{\rm KY}$
are interpreted as spectral functionals.

\paragraph{Discussion.}
The log-determinant construction brings tangent growth of all ranks into
one spectral generating function. Its finite-time and stationary MDEs
retain the complete gain-path distribution and the repeated action of
quenched connectivity. Direct propagation, stationary full-spectrum
comparisons, and recovery of the known maximal exponent provide
complementary tests of this framework. The replay experiment identifies
a physical consequence: temporal memory can redistribute instability
among tangent directions while leaving instantaneous Jacobian statistics
and total contraction unchanged. Spectral stability therefore carries
information about dynamical history beyond the one-time distribution of
local linear responses.

The numerical comparisons assess the theory at finite size, observation
time, and path resolution; their accuracy need not be uniform in these
parameters. The autonomous network's near-onset spectral density and
critical amplitudes remain open questions. Establishing a uniform
interchange of large size, long time, and vanishing regularizer likewise
requires control beyond the present comparisons. The explicit
assumptions and numerical controls delineate these questions while
supporting the spectral description in the regimes examined here.
The appendices give the derivations, numerical protocols, and uncertainty
estimates needed to assess the reported comparisons.

\begin{acknowledgments}
ChatGPT 5.6 Sol assisted with theoretical analysis, numerical implementation and
checking, and manuscript drafting. It is acknowledged here and in the
abstract rather than listed as an author. Scientific claims in this
manuscript are tied to the calculations, saved numerical records, and
explicit assumptions described below; responsibility for the submitted
work rests with the human authors.
\end{acknowledgments}
\appendix
\setcounter{secnumdepth}{2}
\section{Spectral identities and activity input}
\label{app:identities}

\subsection{Activity covariance and the gain law}
The standard asymmetric-network cavity construction gives an effective
single-site process
\begin{equation}
 \dot x(t)=-x(t)+\zeta(t),\qquad
 \E[\zeta(t)\zeta(s)]=g^2 C_\phi(t,s),
 \label{eq:single-site}
\end{equation}
where $C_\phi(t,s)=\E[\tanh x(t)\tanh x(s)]$.
There is no leading retarded self-interaction for the fully asymmetric
independent-coupling ensemble. Applying the linear filter in
Eq.~\eqref{eq:single-site} gives
$\widehat C_x(\omega)=g^2\widehat C_\phi(\omega)/(1+\omega^2)$,
equivalent to Eq.~\eqref{eq:activity}. We use this established activity
DMFT as input \cite{SCS1988,CrisantiSompolinsky2018}.

For completeness, both needed nonlinear two-time correlations can be
computed by ordinary Gaussian quadrature. Let $q_x=C_x(0)$ and
$C=C_x(\tau)$. For independent standard normals $z_1,z_2$, define
\begin{equation}
 X=\sqrt{q_x}z_1,\qquad
 Y=\frac{C}{\sqrt{q_x}}z_1+
 \sqrt{q_x-\frac{C^2}{q_x}}z_2. \label{eq:gaussian-pair}
\end{equation}
Then $C_\phi(\tau)=\E[\tanh X\tanh Y]$ and
$C_d(\tau)=\E[\sech^2X\sech^2Y]$. The chaotic branch has
$C_x'(0)=0$ and $C_x(\tau)\to0$ at large $|\tau|$.
After solving for $C_x$, finite Gaussian vectors with this covariance
are transformed componentwise by $\sech^2$. Keeping these vectors
as entire paths preserves their higher joint gain statistics.
Keeping only $C_d$ would not specify the nonlinear marked inverse.

\subsection{Free energy, smoothing, and exterior powers}
Writing the eigenvalues of $M_T^\dagger M_T$ as $s_i^2$ immediately gives
\begin{equation}
 F_{N,T}(\mu)=\frac1N\sum_i
 \frac{\log[1+e^{2T(\lambda_i^{(T)}-\mu)}]}{2T}.
 \label{eq:softplus}
\end{equation}
Differentiation yields Eq.~\eqref{eq:finite-tail} and
\begin{equation}
 F_{N,T}''(\mu)=\frac{2T}{N}\sum_i p_i(\mu)[1-p_i(\mu)]\geq0,
 \label{eq:convexity}
\end{equation}
where $p_i$ is the logistic factor. Thus $F_{N,T}$ is decreasing and
convex, and its negative derivative lies between zero and one.
The elementary bound
\begin{equation}
 0\leq\frac{\log(1+e^{2Ty})}{2T}-y_+
 \leq\frac{\log2}{2T} \label{eq:softplus-bound}
\end{equation}
separates smoothing error from convergence of the singular exponents.
At fixed $N$, convergence of $\lambda_i^{(T)}$ gives
$F_N(\mu)=N^{-1}\sum_i(\lambda_i-\mu)_+$. Passing to a limiting
empirical measure requires first-moment uniform integrability, not merely
weak convergence. Differentiation at an atom is understood through
one-sided derivatives or distributions.

Expanding $\prod_i(1+qs_i^2)$ gives the elementary symmetric polynomials
in $s_i^2$. The singular values of $\wedge^kM_T$ are the products
$s_{i_1}\cdots s_{i_k}$ over strictly increasing index sets. Its
squared Frobenius norm is therefore precisely the coefficient of $q^k$.
This proves Eq.~\eqref{eq:exterior}, including the $k=0$ term equal to
one. It also explains why a single scalar determinant contains growth
rates of tangent volumes of every rank.

\subsection{Positive sums and the Kaplan--Yorke functional}
Let $\lambda(\alpha)$ be a decreasing spectral quantile and
$\Sigma(\alpha)=\int_0^\alpha\lambda(v)\,\dd v$.
The selection of all exponents above a threshold gives the convex duality
\begin{align}
 F(\mu)&=\sup_{0\leq\alpha\leq1}
 [\Sigma(\alpha)-\mu\alpha],\label{eq:dual-one}\\
 \Sigma(\alpha)&=\inf_{\mu\in\mathbb R}
 [F(\mu)+\mu\alpha]. \label{eq:dual-two}
\end{align}
At a continuous interior crossing, $\alpha=\nu(\mu)$ and
$\Sigma(\alpha)=F(\mu)+\mu\nu(\mu)$, proving Eq.~\eqref{eq:ky}.
For a finite spectrum, we instead use the standard interpolated
definition: if $j$ is the last index with
$\sum_{i\leq j}\lambda_i\geq0$,
\begin{equation}
 D_{\rm KY}=j+\frac{\sum_{i\leq j}\lambda_i}{|\lambda_{j+1}|}.
 \label{eq:finite-ky}
\end{equation}
The all-contracting and all-expanding endpoints require their usual
separate conventions. No assertion that this spectral functional equals
an actual attractor dimension is needed for the numerical comparisons.

\subsection{Contraction and exactly solvable checks}
For continuous time,
\begin{equation}
 \partial_t\log|\det M_t|
 =\Tr[-\Id+WD(t)]=-N, \label{eq:liouville}
\end{equation}
because $W_{ii}=0$. Consequently $\det M_T=e^{-NT}$ and the mean of
both the finite-time singular exponents and the asymptotic exponents
is $-1$. This statement also holds for prescribed gains at any replay
rate. It does not hold exactly for the finite-step approximation
in Eq.~\eqref{eq:step}; the latter is checked with its own determinant.

At $g=0$, all continuous exponents equal $-1$, and
$\nu_{N,T}(\mu)=[1+e^{2T(\mu+1)}]^{-1}$.
For a constant scalar gain $d_0$, the continuous generator is
$-\Id+d_0W$. The asymptotic exponents of this fixed matrix are the real
parts of its eigenvalues; polynomial nonnormal transients do not change
their exponential rates. Projecting the circular law onto the real axis
gives, for $R=gd_0>0$,
\begin{equation}
 \rho(\lambda)=\frac{2}{\pi R^2}
 \sqrt{R^2-(\lambda+1)^2},\quad |\lambda+1|<R.
 \label{eq:constant-spectrum}
\end{equation}
The stable zero-activity branch has $d_0=1$. This is an asymptotic
constant-generator check, not a formula for the finite-time singular
values of a nonnormal matrix.

\section{Finite-time contour and covariance closure}
\label{app:finite}

\subsection{Contour determinant and inverse corner}
Set $n_c=2m$ and introduce
\begin{equation}
 C_k=\begin{cases}
 B_k,&0\leq k<m,\\
 B_{n_c-1-k}^\dagger,&m\leq k<n_c.
 \end{cases} \label{eq:contour-factors}
\end{equation}
Then $C_{n_c-1}\cdots C_0=M_T^\dagger M_T$. For $q>0$ let
\begin{equation}
 \mathbb L_q=\begin{pmatrix}
 \Id&0&\cdots&qC_{n_c-1}\\
 -C_0&\Id&\cdots&0\\
 0&-C_1&\ddots&\vdots\\
 \vdots&&\ddots&\Id
 \end{pmatrix}. \label{eq:contour-matrix}
\end{equation}
The last row contains $-C_{n_c-2}$ immediately below the diagonal.
Eliminating the lower block rows leaves the first diagonal block
$\Id+qC_{n_c-1}\cdots C_0$ and identities elsewhere. Therefore
\begin{equation}
 \det\mathbb L_q=\det(\Id+qM_T^\dagger M_T).
 \label{eq:det-proof}
\end{equation}
The same recursion, applied to a right-hand side supported in block
zero, gives
\begin{equation}
 (\mathbb L_q^{-1})_{00}=(\Id+qM_T^\dagger M_T)^{-1}.
 \label{eq:corner-proof}
\end{equation}
These results do not require invertibility of any individual $B_t$.

Let $\alpha=e^{-\mu\delta}=q^{1/n_c}$ and conjugate by
$\mathbb D_\alpha=\diag(\Id,\alpha\Id,\ldots,
\alpha^{n_c-1}\Id)$. In
$\mathbb L_\alpha=\mathbb D_\alpha\mathbb L_q\mathbb D_\alpha^{-1}$,
all subdiagonal factors become $-\alpha C_k$ and the top-right corner
becomes $+\alpha C_{n_c-1}$. The determinant and inverse corner are
unchanged. This gauge avoids placing an exponentially large or small
weight on a single boundary bond. It is an exact similarity transform,
not a change of the spectral observable.

\subsection{Explicit path matrices}
Write $E_{pq}$ for the $n_c\times n_c$ matrix units, with indices
starting at zero. Define
\begin{align}
 Q&=-\sum_{k=0}^{n_c-2}E_{k+1,k}+E_{0,n_c-1},\nonumber\\
 \ell_0&=\Id_{n_c}+\alpha cQ, \label{eq:skeleton}\\
 F_d&=-\alpha b\sum_{t=0}^{m-1}d^tE_{t+1,t},\label{eq:path-forward}\\
 R_d&=-\alpha b\sum_{t=1}^{m-1}d^tE_{n_c-t,n_c-1-t}
       +\alpha b d^0E_{0,n_c-1}. \label{eq:path-return}
\end{align}
In site-major ordering,
\begin{equation}
 (\mathbb L_\alpha)_{ij}=\delta_{ij}\ell_0+
 \frac{g}{\sqrt N}(\xi_{ij}F_{d_j}+\xi_{ji}R_{d_i}).
 \label{eq:site-contour}
\end{equation}
The return term carries $d_i$ because transposition turns $WD$ into
$DW^\mathsf T$. This distinction is essential.

Hermitize by
$\mathbb H=\left(\begin{smallmatrix}0&\mathbb L_\alpha\\
\mathbb L_\alpha^\dagger&0\end{smallmatrix}\right)$ and group the two
sectors at each site. Its internal matrices are
\begin{equation}
 h_0=\begin{pmatrix}0&\ell_0\\\ell_0^\dagger&0\end{pmatrix},
 \qquad A_d=\begin{pmatrix}0&F_d\\R_d^\dagger&0\end{pmatrix}.
 \label{eq:path-hermitization}
\end{equation}
For $i\ne j$,
\begin{equation}
 \mathbb H_{ij}=\frac{g}{\sqrt N}
 (\xi_{ij}A_{d_j}+\xi_{ji}A_{d_i}^\dagger).
 \label{eq:random-block}
\end{equation}
This expression contains a single random variable for all appearances
of the same directed connection on the contour.

\subsection{Gaussian contraction and the Dyson equation}
For a path-conditioned matrix field $R_d$, average the product
$\mathbb H_{ij}R_{d_j}\mathbb H_{ji}$ at fixed path marks.
Since $\E\xi_{ij}^2=1$ and $\E\xi_{ij}\xi_{ji}=0$,
\begin{align}
 \E_\xi[\mathbb H_{ij}R_{d_j}\mathbb H_{ji}]
 =\frac{g^2}{N}\big(&A_{d_j}R_{d_j}A_{d_j}^\dagger\nonumber\\
 &+A_{d_i}^\dagger R_{d_j}A_{d_i}\big).
 \label{eq:contraction}
\end{align}
Summing over $j$ in the independent marked model gives the covariance map
\begin{equation}
 (\mathcal S R)_d=g^2\left[
 \E_{d'}(A_{d'}R_{d'}A_{d'}^\dagger)
 +A_d^\dagger\E_{d'}R_{d'}A_d\right].
 \label{eq:covariance-map}
\end{equation}
It is completely positive, preserves adjoints, and is symmetric for the
trace inner product $\E_d\Tr(R_d^\dagger Q_d)$. For bounded paths,
$\norm{\mathcal S}\leq2g^2\sup_d\norm{A_d}^2$.
The associated Dyson equation is
\begin{equation}
 \mathcal M_d^{-1}=h_0-z\Id-(\mathcal S\mathcal M)_d.
 \label{eq:dyson-general}
\end{equation}
This is the block Gaussian resolvent closure at fixed internal dimension
and positive $\Im z$, with the upper-half-plane branch chosen by
$\mathcal M(z)\sim-z^{-1}\Id$ at large $|z|$. The operator-valued
Dyson framework supplies the natural analytic formulation
\cite{Helton2007}. No microscopic local law or flat covariance assumption
is needed for the numerical use made here.

At $z=0$ for a finite invertible contour,
\begin{equation}
 \mathbb H^{-1}=
 \begin{pmatrix}0&(\mathbb L_\alpha^\dagger)^{-1}\\
 \mathbb L_\alpha^{-1}&0\end{pmatrix}.
 \label{eq:hermitized-inverse}
\end{equation}
The lower-left corner and Eq.~\eqref{eq:corner-proof} give
Eq.~\eqref{eq:finite-readout}. The passage to a large-$N$
zero-regulator boundary value is distinct from the fixed-positive-regulator
Gaussian contraction. The numerical tests explicitly reduce the
regulator and compare with the exact finite-product observable.

\subsection{Autonomous paths and the scope of the closure}
In the autonomous network $D(t)$ depends on $W$. One cannot condition
on the complete gain path and then treat $W$ as an untouched independent
Gaussian matrix. For fixed initial conditions independent of $W$,
differentiating the flow gives the exact sensitivity identity
\begin{equation}
 \partial_{\xi_{ij}}x_k(t)=\frac{g}{\sqrt N}
 \int_0^t\Phi_{ki}(t,s)\tanh x_j(s)\,\dd s,
 \label{eq:sensitivity}
\end{equation}
where $\Phi(t,s)$ is the tangent fundamental matrix. On a bounded
connectivity-norm event, $\norm{\Phi(t,s)}\leq
e^{(1+\norm W)(t-s)}$. This gives finite-horizon sensitivity control,
but its constants grow with the horizon. It does not alone prove that
the sum of every path-derivative term in a Gaussian integration by
parts vanishes at long times.

For this reason the autonomous-network application is stated as a
tested mean-field replacement. The paired finite-time SVD and stationary
QR comparisons in Appendix~\ref{app:numerics} probe that replacement at the reported sizes
and times. We do not promote them, or Eq.~\eqref{eq:sensitivity},
to an unqualified stationary cavity theorem.

\section{Stationary cocycle, operator equation, and edge}
\label{app:stationary}

\subsection{Exact cocycle representation}
Fix $N$ and a realization of $W$. Let
$(\Omega,\mathbb P,\Theta)$ be a two-sided stationary ergodic
environment for the gain paths. Ergodicity is assumed for the actual
joint environment; it need not follow from ergodicity of one factor in
an arbitrary product system. Set $B(\omega)=c\Id+bWD(\omega)$.
The chronological product is
\begin{equation}
 M_m(\omega)=B(\Theta^{m-1}\omega)\cdots B(\omega).
 \label{eq:chronological}
\end{equation}
Choose a shift unitary with
$UfU^*=f\circ\Theta^{-1}$ and put $Y=BU$. Then
\begin{equation}
 Y^m=B(\omega)B(\Theta^{-1}\omega)\cdots
 B(\Theta^{-(m-1)}\omega)U^m. \label{eq:cocycle-power}
\end{equation}
Changing the time anchor by stationarity recovers the matrix order in
Eq.~\eqref{eq:chronological}. Neither the connectivity nor the time
ordering has been replaced.

Use normalized trace $\tau_N=N^{-1}\Tr_N\otimes\tau_{\rm erg}$
on the matrix crossed-product algebra. If $B$ is an essentially bounded
measurable invertible cocycle with
$\E\log^+\norm{B^{-1}}<\infty$, Deninger's theorem
\cite{Deninger2011} gives
\begin{equation}
 \log\Delta_{\tau_N}(\Id-Y)=\frac1N\sum_i(\chi_i)_+,
 \label{eq:deninger}
\end{equation}
where $\chi_i=\delta\lambda_i$ are the per-step exponents.
Scaling $Y$ by $e^{-\delta\mu}$ and using
$\tau_N(\Id)=1$ proves Eq.~\eqref{eq:brown-free}.
The theorem is used under its integrability conditions; invertibility
almost surely alone does not establish integrability of the inverse.

The gauge rotation $U\mapsto e^{\ii\theta}U$ preserves the trace.
Thus the Brown logarithmic potential
$\mathcal U(r)=\log\Delta_\tau(r\Id-Y)$ is radial. Applying
Eq.~\eqref{eq:deninger} to every positive scalar multiple gives
\begin{equation}
 \mathcal U(r)=\frac1N\sum_i\max(\log r,\chi_i).
 \label{eq:radial-potential}
\end{equation}
It is the potential of mass $1/N$ uniformly distributed on each circle
of radius $e^{\chi_i}$. Consequently, away from atoms,
\begin{equation}
 r\mathcal U'(r)=\mu_Y(\{|z|\leq r\}),\qquad
 \nu(\mu)=1-r\mathcal U'(r). \label{eq:radial-cdf}
\end{equation}
This representation is exact for the fixed-$N$ cocycle under the
stated hypotheses. It does not require nor imply the MDE approximation.

\subsection{Stationary Hermitization and covariance}
On the time lattice, the cocycle pencil has the form
\begin{equation}
 \begin{gathered}
 (r\Id-Y)_{ij}=\delta_{ij}\ell_r+
 \frac{g}{\sqrt N}\xi_{ij}K_{d_j},\\
 \ell_r=r\Id-cS,\quad K_d=-bD_dS.
 \end{gathered}
 \label{eq:stationary-pencil}
\end{equation}
The choice of gain index on $D_dS$ is a stationary anchoring convention.
Hermitizing produces
\begin{equation}
 h_r=\begin{pmatrix}0&\ell_r\\\ell_r^\dagger&0\end{pmatrix},\qquad
 \mathcal A_d=\begin{pmatrix}0&K_d\\0&0\end{pmatrix}.
 \label{eq:stationary-A}
\end{equation}
The two contractions are exactly those of
Eq.~\eqref{eq:covariance-map}, with $A_d$ replaced by
$\mathcal A_d$. In particular,
\begin{align}
 \mathcal A_d\mathcal M_d\mathcal A_d^\dagger
 &=\begin{pmatrix}K_dG_{--,d}K_d^\dagger&0\\0&0\end{pmatrix},\nonumber\\
 \mathcal A_d^\dagger\overline{\mathcal M}\mathcal A_d
 &=\begin{pmatrix}0&0\\0&K_d^\dagger P K_d\end{pmatrix}.
 \label{eq:stationary-contractions}
\end{align}
This yields Eqs.~\eqref{eq:stationary-mde} and
\eqref{eq:stationary-orders}. Stationarity of the path law ensures
$SPS^\dagger=P$ and $S\Gamma S^\dagger=\Gamma$, explaining the
two Fourier order parameters.

For a finite time volume, let $L_r=r\Id-Y$ and define the regularized
potential $\mathcal U_\eta=(2NL)^{-1}
\log\det(L_r^\dagger L_r+\eta^2\Id)$.
Differentiation gives
\begin{equation}
 \partial_r\mathcal U_\eta
 =\frac1{NL}\Re\Tr[(L_r^\dagger L_r+\eta^2\Id)^{-1}L_r^\dagger].
 \label{eq:regularized-potential}
\end{equation}
The lower-left block of
$[\left(\begin{smallmatrix}0&L_r\\L_r^\dagger&0\end{smallmatrix}\right)
-\ii\eta\Id]^{-1}$ is the matrix inside this trace.
Equation~\eqref{eq:stationary-readout} follows after normalized averaging
and the boundary limit. A finite positive-$\eta$ readout need not lie
in $[0,1]$; only a legitimate limiting cumulative spectrum must do so.

\subsection{Limits and assumptions}
At each fixed time volume, positive regulator, and independent bounded
path law, the problem is a Gaussian block resolvent problem. Enlarging
the time volume additionally requires the periodic path populations to
approach the intended stationary law in finite windows. Fixed-positive-
regulator boundary estimates are not uniform as $\eta\downarrow0$.
Convergence of logarithmic potentials needs control of the small
singular values, for example uniform integrability of $|\log s|$ near
zero. The numerical sequence is therefore organized by separate
population, volume, regularizer, and step controls, not by an arbitrary
joint scaling of these parameters.

For the autonomous network a further assumption is that the stationary
neural gain histories can be replaced, in this normalized spectral
observable, by independent histories with the activity-DMFT law.
We do not supply a uniform proof of that replacement here. Its
empirical support in the present work is the explicit full-curve and
paired-cocycle comparison. This distinction remains even though the
stationary representation, the Gaussian covariance contractions, and
the maximal-exponent reduction each have exact algebraic content.

\subsection{Exterior stability and the two-copy equation}
\label{app:edge-reduction}
At $\eta=0$ on the exterior branch, $P=\Gamma=0$: the diagonal
Green blocks vanish, while the off-diagonal blocks are determined by
$\ell_r^{-1}$ and $(\ell_r^\dagger)^{-1}$. Linearizing the diagonal
sector of the same stationary MDE gives the positive map
\begin{equation}
 \mathcal T_r[Q]=g^2\E_d[(\ell_r^\dagger)^{-1}
 K_d^\dagger QK_d\ell_r^{-1}]. \label{eq:edge-map}
\end{equation}
Each term has the form $A_d^\dagger Q A_d$, with
$A_d=K_d\ell_r^{-1}$, so positive semidefinite perturbations remain
positive semidefinite. Starting from the stable large-$r$ exterior
branch, the outer boundary occurs when the spectral radius of
$\mathcal T_r$ reaches one. This is a linear stability condition on
the full-spectrum equation.

Stationarity allows $Q_{ts}=q_{t-s}$. Define the uncentered gain
correlation $C_d(\tau)=\E[d(0)d(\tau)]$; its nonzero long-lag
limit must not be subtracted. With $K_d=-bD_dS$, stationarity gives
\begin{equation}
 [\E K_d^\dagger QK_d]_{ts}
 =b^2C_d(t-s)q_{t-s}. \label{eq:lag-product}
\end{equation}
At a unit eigenvalue of the stability map, multiplication by
$\ell_r^\dagger$ and $\ell_r$ gives
$\ell_r^\dagger Q\ell_r=g^2\E K_d^\dagger QK_d$.
Fourier transformation back to the lag variable yields
\begin{equation}
 \big[(r^2+c^2)\Id-rc(S_\tau+S_\tau^\dagger)
       -g^2b^2\diag C_d\big]q=0, \label{eq:discrete-edge}
\end{equation}
where $S_\tau$ now shifts the lag coordinate. The first instability
occurs when this Hermitian operator has zero lowest eigenvalue.

For $r=e^{\delta\lambda}$ and $c=e^{-\delta}$, its Fourier kinetic
symbol is
\begin{equation}
 \delta^{-2}[r^2+c^2-2rc\cos(\omega\delta)]
 \longrightarrow(1+\lambda)^2+\omega^2.
 \label{eq:kinetic-limit}
\end{equation}
Also $b/\delta\to1$. Dividing Eq.~\eqref{eq:discrete-edge} by
$\delta^2$ therefore gives the continuous-lag equation
\begin{equation}
 [-\partial_\tau^2+(1+\lambda)^2-g^2C_d(\tau)]\psi(\tau)=0.
 \label{eq:edge}
\end{equation}
Introduce the self-adjoint one-dimensional operator
\begin{equation}
 \mathcal H=-\partial_\tau^2+V_{\rm eff}(\tau),\qquad
 V_{\rm eff}(\tau)=1-g^2C_d(\tau).
 \label{eq:edge-hamiltonian}
\end{equation}
For the stationary bounded gains considered here, the potential is
real, even, and bounded. The infinite-lag operator acts on
$L^2(\mathbb R)$ with domain $H^2(\mathbb R)$; finite-volume
periodic approximations and their controls are described in
Appendix~\ref{app:edge-numerics}. Let $E_0=\inf\sigma(\mathcal H)$,
the ground-state energy when the bottom of the spectrum is a discrete
eigenvalue. Equation~\eqref{eq:edge} then gives
\begin{equation}
 E_0+(1+\lambda_e)^2-1=0.
 \label{eq:edge-ground-energy}
\end{equation}
The first instability reached from large positive $\lambda$ selects
the root $1+\lambda_e>0$, yielding Eq.~\eqref{eq:edge-value}.
The time-lag coordinate and the gain correlation thus define an
effective spectral potential whose lowest energy controls asymptotic
tangent growth. The Schr\"odinger form is a representation of this
classical stability problem, without introducing quantum dynamics.

This is precisely the maximal-exponent problem of two-copy DMFT
\cite{SCS1988,CrisantiSompolinsky2018}. Only two gain factors occur
in the linear contraction in Eq.~\eqref{eq:lag-product}, which explains
the reduction to $C_d$. Inside the support, the gain-dependent inverse
in the nonlinear stationary MDE still requires the full path law.
The reduction identifies the outer boundary; it does not determine a
local spectral-density shape there or establish the same reduction for
arbitrary nonlinear random networks. No extrapolation from a small
positive tail threshold enters this calculation.

\subsection{Schur implementation and regularizer response}
The implemented stationary solver uses a periodic time shift on $L$
sites and $R$ sampled paths. Let
\begin{align}
 A&=-z\Id-\Gamma,\qquad R_A=A^{-1},\nonumber\\
 Q_d&=-z\Id-g^2K_d^\dagger P K_d-\ell_r^\dagger R_A\ell_r.
 \label{eq:schur-Q}
\end{align}
Only $L\times L$ path matrices need inversion:
\begin{align}
 G_{--,d}&=Q_d^{-1},\nonumber\\
 G_{++,d}&=R_A+R_A\ell_rQ_d^{-1}\ell_r^\dagger R_A,\nonumber\\
 G_{-+,d}&=-Q_d^{-1}\ell_r^\dagger R_A. \label{eq:schur-blocks}
\end{align}
Path averaging and projection onto the Fourier diagonal update $P$
and $\Gamma$. The projection enforces the stationary translation
average for a finite sampled population. Damping and continuation from
large to small $\eta$ select the resolvent branch. The readout uses the
exact normalized trace at the numerical cutoff.

For the derivative correction, collect the two Fourier symbols in $v$
and write $v=\Psi(v,\eta)$. At a nonsingular linearized fixed point,
\begin{equation}
 (\Id-D_v\Psi)\dot v=\partial_\eta\Psi,
 \qquad \dot v=\partial_\eta v. \label{eq:implicit}
\end{equation}
This derivative is computed by differentiating the inverses, not by
fitting a coefficient to QR data. Explicitly,
\begin{align}
 \dot A&=-\ii\Id-\dot\Gamma,\qquad
 \dot R_A=-R_A\dot A R_A,\nonumber\\
 \dot Q_d&=-\ii\Id-g^2K_d^\dagger\dot P K_d
            -\ell_r^\dagger\dot R_A\ell_r,\nonumber\\
 \dot G_{--,d}&=-G_{--,d}\dot Q_dG_{--,d},\label{eq:schur-response}\\
 \dot G_{-+,d}&=-\dot G_{--,d}\ell_r^\dagger R_A
                -G_{--,d}\ell_r^\dagger\dot R_A.\nonumber
\end{align}
Differentiating $G_{++}$ in Eq.~\eqref{eq:schur-blocks} and closing
$\dot P=\Pi_{\rm circ}\E\dot G_{++}$ and
$\dot\Gamma=g^2\Pi_{\rm circ}\E(K\dot G_{--}K^\dagger)$
completes the affine linear response problem. Its readout is
\begin{equation}
 \partial_\eta\nu_{L,\eta}
 =-\frac rL\Re\Tr\E\dot G_{-+,d}. \label{eq:eta-readout}
\end{equation}
If at the threshold considered
$\nu_\eta=\nu_0+A_1\eta+A_2\eta^2+O(\eta^3)$, then
\begin{align}
 t(\eta)&=\nu_\eta-\eta\partial_\eta\nu_\eta
          =\nu_0-A_2\eta^2+O(\eta^3),\nonumber\\
 \widehat\nu_0&=\frac{q_\eta^2t(e)-t(q_\eta e)}{q_\eta^2-1},
 \qquad q_\eta>1. \label{eq:richardson}
\end{align}
The remainder diagnostic is $|\widehat\nu_0-t(e)|$.
The derivative equation is exact at fixed positive regulator and
cutoffs. The extrapolation additionally assumes the displayed regular
expansion and can lose validity at a singular edge; the exterior
stability problem is used there instead.

\section{Numerical validation and uncertainty}
\label{app:numerics}

Finite-time comparisons use the logistic tail of singular values;
long-time QR comparisons use the hard cumulative tail. Direct stationary
matrices and their MDE are compared at the same positive regulator.
Independent connectivity matrices and independent parent path populations
are the respective sampling units. Thresholds, translated windows, and
time prefixes from one realization do not increase the sample count.
Numerical cutoff envelopes describe systematic uncertainty, not
statistical confidence intervals.

\subsection{Finite-time validation}
The short-time test uses $g=1.5$, $T=0.5$, $\delta=0.05$, and
$N=128,256,512,1024$, with constant gains, independent gain marks,
and correlated DMFT paths. Each direct product retains one fixed matrix
throughout, with eight realizations per mean curve. MDE residuals are
below $5\times10^{-10}$; the checked Herglotz eigenvalues are positive,
and the final regulator change is below $7.5\times10^{-5}$.

The autonomous-network comparison uses $\delta=0.05$ and
$N=256,512,1024$ with $3,3,2$ matrices for $g=1.2,3$;
for $g=1.5$, the sizes $256,512,1024,2048$ use $4,4,3,1$ matrices.
Heun integration discards transients $120,100,80$ for $g=1.2,1.5,3$,
respectively; both tangent systems use Eq.~\eqref{eq:step}.
The six horizons share product prefixes. On 25 equally spaced thresholds
in $[-2.8,1.2]$, the MDE path counts at successive horizons are
$(128,64,64,16,16,4)$ for $g=1.2,3$ and
$(256,128,256,64,16,4)$ for $g=1.5$; regulator continuation ends
at $3\times10^{-4}$. A $T=2$, $g=1.5$, $N=512$ control adds
steps $0.1,0.025,0.0125$, with three matrices per step.
Varying path resolution prevents attributing the long-horizon errors
to observation time alone.

\subsection{Stationary closure and network comparison}
\begin{figure*}[t]
\centering\includegraphics[width=.95\textwidth]{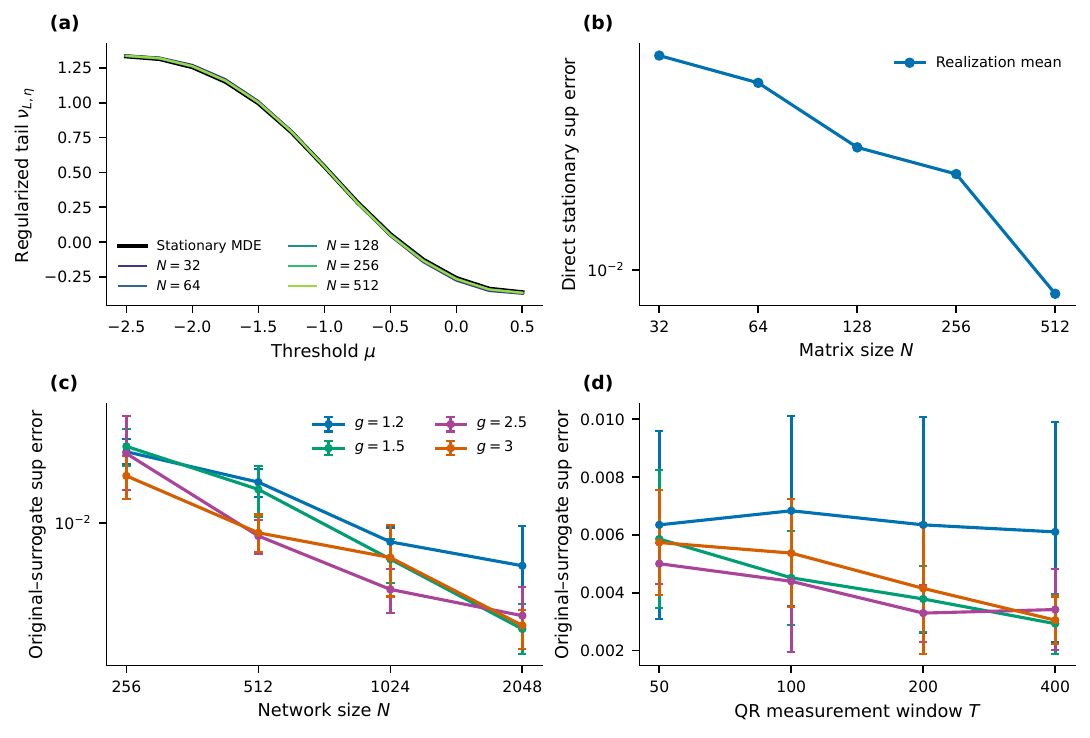}
\caption{Stationary closure and path-replacement tests.
(a),(b) Direct regularized matrices versus the matching MDE at
$(L,\eta,\delta)=(12,0.03,0.05)$; (b) shows mean realization-level
supremum errors. (c) Paired network/path hard-tail discrepancies with
bootstrap intervals. (d) Prefix comparisons from the same $N=2048$
QR histories.}
\label{fig:closure}
\end{figure*}

Direct Hermitized matrices are compared with the stationary MDE at
$(L,\eta,\delta)=(12,0.03,0.05)$. Sizes
$N=32,64,128,256,512$ use $16,16,12,8,4$ realizations.
The mean realization-level supremum error decreases with size;
the error of the mean curve fluctuates and need not decrease
monotonically [Fig.~\ref{fig:closure}(a),(b)].

The paired network/path test uses $N=256,512,1024,2048$ with
$16,12,8,4$ matrices, the same connectivity and discretization in
each pair, and full-rank QR over $T=400$ \cite{Benettin1980}.
Paired tail discrepancies decrease with size; prefixes at
$T=50,100,200,400$ show no persistent increase at $N=2048$
[Fig.~\ref{fig:closure}(c),(d)]. These tests do not resolve spectral
mass much smaller than $1/N$ or imply convergence uniform in time.

The QR reference at each threshold is the median of the $N=2048$
endpoint, a $1/N$ intercept, and a $1/\sqrt N$ intercept.
A combined $1/\sqrt N+1/N$ fit contributes to the model envelope.
Connectivity bootstraps give sampling bands; size-model spread and
measurement-time variation are assessed separately.

\subsection{Numerical cutoffs and uncertainty budget}
\begin{figure*}[t]
\centering\includegraphics[width=.95\textwidth]{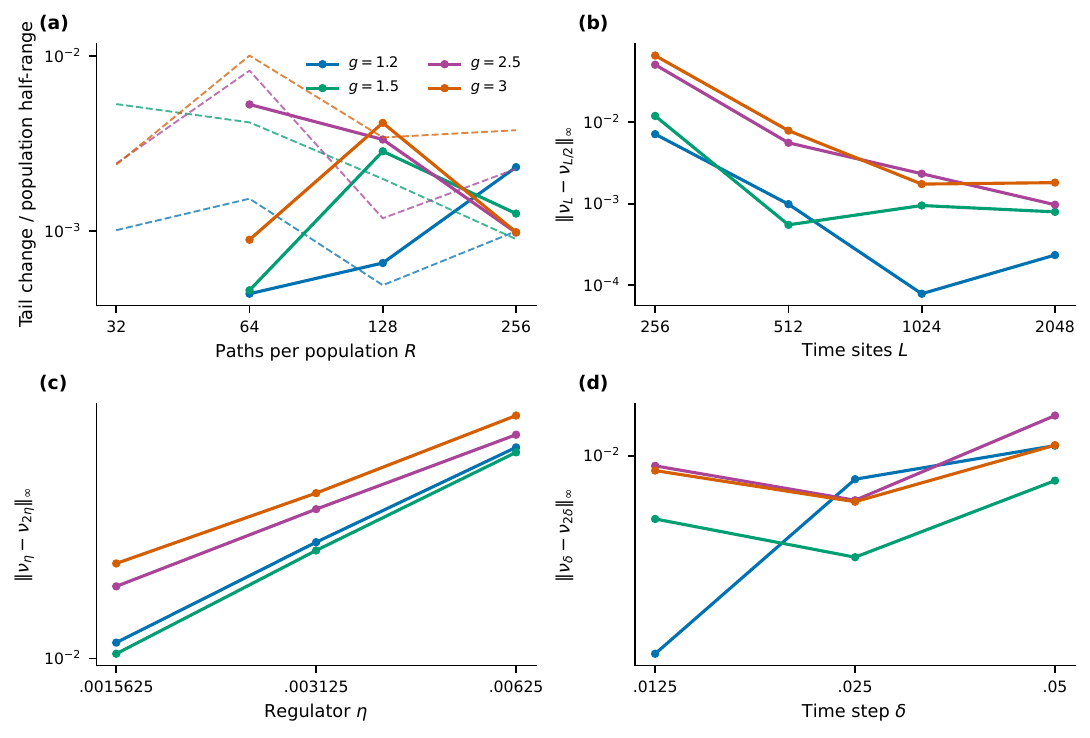}
\caption{Sensitivity to the numerical cutoffs. (a) Successive path-count
changes (solid) and the half-range of four populations (dashed).
(b) Successive time-volume changes at the smallest regulator.
(c) Successive regulator changes at $L=2048$.
(d) Successive time-step changes after volume and regulator
extrapolation at each step. Each change is a supremum over the
threshold grid; the abscissa gives the finer cutoff. These are
refinement diagnostics, not bounds on the remaining limiting error.}
\label{fig:controls}
\end{figure*}

The principal MDE spectra use four independent populations of $R=256$
paths, $L=512$ time sites, and step $\delta=0.05$.
Here $R$ controls path sampling, $L\delta$ the time volume,
$\eta$ the Hermitization regulator, and $\delta$ the discretization.
Controls vary $R$ from 32 to 256 and $L$ from 128 to 2048, and decrease $\eta$
to $0.0015625$ and $\delta$ to $0.0125$ (Fig.~\ref{fig:controls}).
Step comparisons keep physical periods fixed; translated volume
windows share their parent history. Extrapolated tails are clipped to
$[0,1]$ and made monotone without using QR targets. Integrated
observables use piecewise cubic Hermite interpolation.

MDE residuals are below $3.1\times10^{-9}$, and the checked Herglotz
eigenvalues exceed $9.7\times10^{-5}$. Full positivity checks extend
through $L=1024$; $L=2048$ follows the continued branch.
The largest last-step tail changes are $0.00182$ for
$L=1024\to2048$ and $0.00923$ for $\delta=0.025\to0.0125$.
These diagnostics do not replace the extrapolation envelopes in
Table~\ref{tab:budgets}. Regulator and time-volume extrapolations
dominate the MDE uncertainty, followed by the time step.
The resulting budgets support full-curve compatibility, not uniform
relative precision in sparse tails. Direct-matrix and paired-path
discrepancies are separate tests and are not added again to the budget.
The $g=2.5$ prediction and reconstruction rules were fixed before its
40 QR and 40 paired-path reference realizations.

\begin{table*}[t]
\caption{Full-spectrum sup-norm comparison. The total sums the MDE
radius, QR sampling radius, size-model spread, and time variation.
Systematic envelopes are not confidence bands. The dedicated
prediction-test time radius is used for $g=2.5$.}
\label{tab:budgets}
\begin{ruledtabular}\begin{tabular}{crrrrrr}
$g$ & observed error & MDE & QR sampling & QR size & QR time & total\\
1.2&0.007219&0.036387&0.005279&0.009949&0.003830&0.055446\\
1.5&0.006825&0.050043&0.004263&0.027902&0.004364&0.086571\\
2.5&0.016483&0.079811&0.002995&0.002785&0.003418&0.089009\\
3.0&0.017768&0.117790&0.003140&0.004433&0.006027&0.131390\\
\end{tabular}\end{ruledtabular}
\end{table*}

\subsection{Positive-spectrum precision}
\begin{figure*}[t]
\centering\includegraphics[width=.94\textwidth]{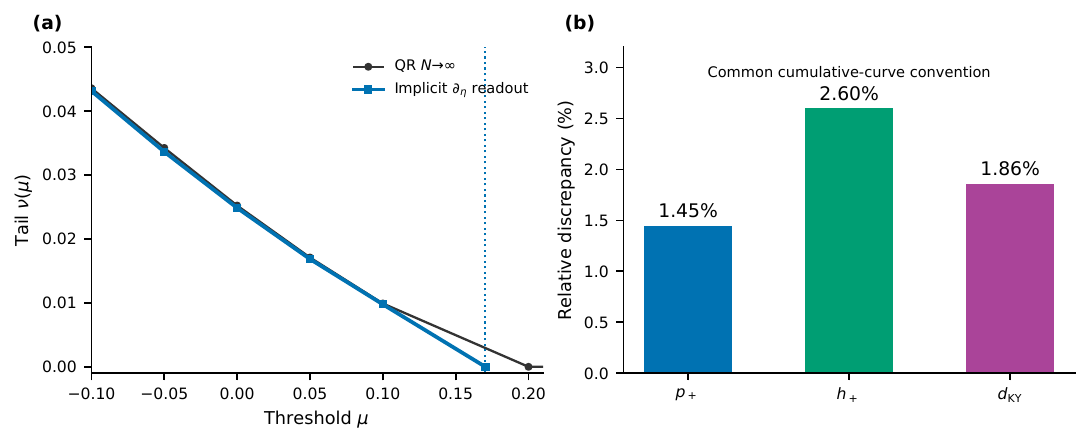}
\caption{Positive-spectrum precision at $g=2.5$ with
$(L,\delta,R)=(512,0.05,256)$. (a) Implicit-derivative readout and
QR tail; shading gives the range of four populations.
(b) Relative discrepancies in $p_+$, $h_+$, and $d_{\rm KY}$ under
a common cumulative-curve convention.}
\label{fig:positive}
\end{figure*}

At $g=2.5$, Eq.~\eqref{eq:richardson} uses four $R=256$
populations at $(L,\delta)=(512,0.05)$, with eight thresholds in
$[-0.3,0.1]$ and support closed at the independently computed
same-step edge. Under a common cumulative-curve convention, the
relative MDE--QR differences in $p_+$, $h_+$, and $d_{\rm KY}$ are
about $1\%$--$3\%$ (Fig.~\ref{fig:positive}).
The near-zero tail sup discrepancy is $9.62\times10^{-4}$.
This test used fixed cutoffs and was performed after the QR reference
was known; it does not repeat the full step and volume extrapolation.
The threshold grid and support interpolation further limit the
precision claim.

\subsection{Maximal-exponent consistency}
\label{app:edge-numerics}
\begin{figure*}[t]
\centering\includegraphics[width=.95\textwidth]{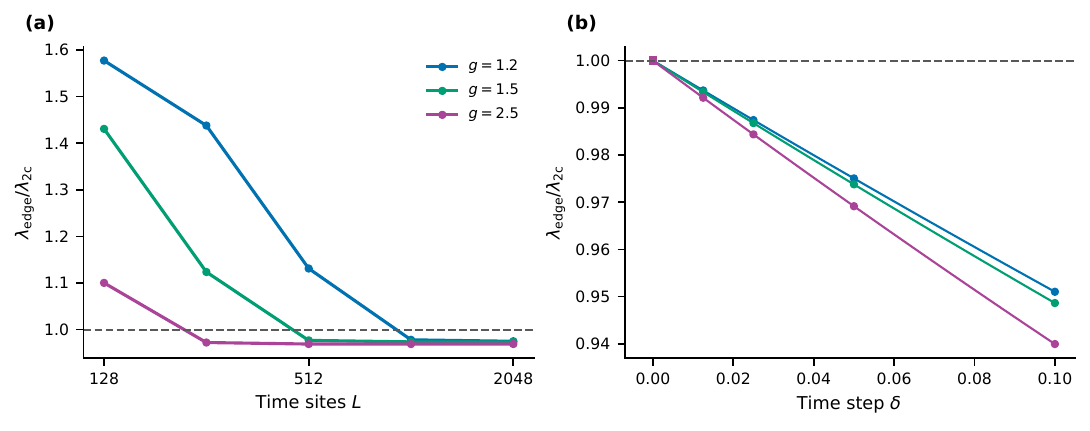}
\caption{Exterior-edge consistency controls. (a) Time-volume refinement.
(b) Step refinement at fixed physical period $102.4$; curves are
quadratic fits, and squares mark their continuous-time intercepts.
The dashed reference is the independently computed two-copy result.}
\label{fig:edge-check}
\end{figure*}

Equation~\eqref{eq:discrete-edge} is extrapolated quadratically in
$\delta$, using four steps obtained by halving from $0.1$ through
$0.0125$ at physical period $102.4$. An independent positive-map
application gives Perron-radius residuals below $9.1\times10^{-12}$.
The two-copy implementation uses the same activity-DMFT input;
the numerical agreement reported in the main text tests the reduction
and its discretization. Figure~\ref{fig:edge-check} shows volume and
step controls. Near-critical autonomous networks have substantial
finite-size effects; this consistency check does not establish a
high-precision direct network-edge measurement.

\section{Temporal replay: validation and limits}
\label{app:replay}

\subsection{Controlled gain histories}
The replay model uses Gaussian stationary histories independently
across sites and independently of $W$, with
$d_a(t)=\sech^2x(at)$ and a fixed leak coefficient.
Activity DMFT supplies the covariance; periodic Gaussian sampling
preserves its variance, without fitting Lyapunov observables.
Stationarity preserves the instantaneous Jacobian distribution across
replay rates, while multitime statistics change.
Equation~\eqref{eq:liouville} fixes the mean exponent at $-1$.
This intervention acts on prescribed gain histories, not on an
autonomous neural attractor.

\subsection{Finite-time comparison}
At $g=1.3$, independent direct products and the MDE use the same
exponential-Euler step, with 12 connectivity matrices per group.
The comparisons cover $a=1,3,8$, $T=8,16$,
$\mu=0,0.08,0.16$, and $N=512,1024$ at $\delta=0.25$;
$\delta=0.125$ is also tested at $N=1024,T=16$.
MDE populations use $R=512,2048$ at the base step and $R=1024$
at the smaller step. Positive path weights match the analytic gain
mean and second moment; direct histories are unweighted and independent
of the MDE population.
The 45 comparisons from 15 parameter groups are compatible with
pointwise 95\% Student intervals for the direct means.
Shared samples make these comparisons correlated, and the intervals
cover matrix sampling only. The largest absolute relative deviation,
normalized by the MDE prediction, is $23.15\%$ in the thin tail;
sampling compatibility therefore does not imply uniform relative
accuracy.

\subsection{Paired long-time intervention}
RK4--QR propagation uses four independent matrices at each of
$N=2048,4096$, paired across replay rates, with 48 tangent directions.
Alignment time 500 is discarded before measurement over time 1500;
the base step is $0.1$, with gain interpolation at source step $0.025$.
All reported counts and thresholds lie within the retained directions.
The mean reductions from $a=1$ to $a=3$ are $3.75$ and $7.50$,
with pointwise 95\% Student intervals $[2.23,5.27]$ and $[6.58,8.42]$,
respectively, computed from the four within-matrix differences.
These are small-sample estimates of the intervention; positive-count
outliers persist at $a=8$. Halving the step at $N=2048$ changes
any retained exponent by at most $1.92\times10^{-5}$ and preserves
all positive counts. This integration check does not remove
finite-size or finite-measurement-time bias.

\begin{table}[!htbp]
\caption{The two stationary replay deviations outside pointwise 95\%
sampling intervals, at $g=1.3$, $N=4096$, and $a=1$.}
\label{tab:replay-absolute}
\begin{ruledtabular}\begin{tabular}{rrrr}
$\mu$ & MDE count & QR mean & 95\% interval\\
$-0.02$ & 24.0368 & 22.75 & $[21.9544,23.5456]$\\
$-0.00456753$ & 13.8816 & 12.50 & $[11.5813,13.4187]$\\
\end{tabular}\end{ruledtabular}
\end{table}

\subsection{Limits of the absolute stationary prediction}
Two of the 30 stationary replay comparisons exceed their pointwise
sampling intervals, both at $N=4096,a=1$ (Table~\ref{tab:replay-absolute}).
Increasing the MDE population from $R=512$
to 2048 changes the first prediction to about $24.02$ and does not
remove its discrepancy. These intervals include matrix sampling only,
without a complete finite-size, finite-time, or MDE numerical error
budget. The differences limit absolute-spectrum precision; they do
not invalidate the directly measured memory effect or establish
failure of the limiting theory.


\begin{thebibliography}{12}\makeatletter
\providecommand \@ifxundefined [1]{\@ifx{#1\undefined}
}\providecommand \@ifnum [1]{\ifnum #1\expandafter \@firstoftwo
 \else \expandafter \@secondoftwo
 \fi
}\providecommand \@ifx [1]{\ifx #1\expandafter \@firstoftwo
 \else \expandafter \@secondoftwo
 \fi
}\providecommand \natexlab [1]{#1}\providecommand \enquote  [1]{``#1''}\providecommand \bibnamefont  [1]{#1}\providecommand \bibfnamefont [1]{#1}\providecommand \citenamefont [1]{#1}\providecommand \href@noop [0]{\@secondoftwo}\providecommand \href [0]{\begingroup \@sanitize@url \@href}\providecommand \@href[1]{\@@startlink{#1}\@@href}\providecommand \@@href[1]{\endgroup#1\@@endlink}\providecommand \@sanitize@url [0]{\catcode `\\12\catcode `\$12\catcode
  `\&12\catcode `\#12\catcode `\^12\catcode `\_12\catcode `\%12\relax}\providecommand \@@startlink[1]{}\providecommand \@@endlink[0]{}\providecommand \url  [0]{\begingroup\@sanitize@url \@url }\providecommand \@url [1]{\endgroup\@href {#1}{\urlprefix }}\providecommand \urlprefix  [0]{URL }\providecommand \Eprint [0]{\href }\providecommand \doibase [0]{https://doi.org/}\providecommand \selectlanguage [0]{\@gobble}\providecommand \bibinfo  [0]{\@secondoftwo}\providecommand \bibfield  [0]{\@secondoftwo}\providecommand \translation [1]{[#1]}\providecommand \BibitemOpen [0]{}\providecommand \bibitemStop [0]{}\providecommand \bibitemNoStop [0]{.\EOS\space}\providecommand \EOS [0]{\spacefactor3000\relax}\providecommand \BibitemShut  [1]{\csname bibitem#1\endcsname}\let\auto@bib@innerbib\@empty
\bibitem [{\citenamefont {Sompolinsky}\ \emph {et~al.}(1988)\citenamefont
  {Sompolinsky}, \citenamefont {Crisanti},\ and\ \citenamefont
  {Sommers}}]{SCS1988}\BibitemOpen
  \bibfield  {author} {\bibinfo {author} {\bibfnamefont {H.}~\bibnamefont
  {Sompolinsky}}, \bibinfo {author} {\bibfnamefont {A.}~\bibnamefont
  {Crisanti}},\ and\ \bibinfo {author} {\bibfnamefont {H.-J.}\ \bibnamefont
  {Sommers}},\ }\href {https://doi.org/10.1103/PhysRevLett.61.259} {\bibfield
  {journal} {\bibinfo  {journal} {Phys. Rev. Lett.}\ }\textbf {\bibinfo
  {volume} {61}},\ \bibinfo {pages} {259} (\bibinfo {year} {1988})}\BibitemShut
  {NoStop}\bibitem [{\citenamefont {Crisanti}\ and\ \citenamefont
  {Sompolinsky}(2018)}]{CrisantiSompolinsky2018}\BibitemOpen
  \bibfield  {author} {\bibinfo {author} {\bibfnamefont {A.}~\bibnamefont
  {Crisanti}}\ and\ \bibinfo {author} {\bibfnamefont {H.}~\bibnamefont
  {Sompolinsky}},\ }\href {https://doi.org/10.1103/PhysRevE.98.062120}
  {\bibfield  {journal} {\bibinfo  {journal} {Phys. Rev. E}\ }\textbf {\bibinfo
  {volume} {98}},\ \bibinfo {pages} {062120} (\bibinfo {year}
  {2018})}\BibitemShut {NoStop}\bibitem [{\citenamefont {Engelken}\ \emph {et~al.}(2023)\citenamefont
  {Engelken}, \citenamefont {Wolf},\ and\ \citenamefont
  {Abbott}}]{Engelken2023}\BibitemOpen
  \bibfield  {author} {\bibinfo {author} {\bibfnamefont {R.}~\bibnamefont
  {Engelken}}, \bibinfo {author} {\bibfnamefont {F.}~\bibnamefont {Wolf}},\
  and\ \bibinfo {author} {\bibfnamefont {L.~F.}\ \bibnamefont {Abbott}},\
  }\href {https://doi.org/10.1103/PhysRevResearch.5.043044} {\bibfield
  {journal} {\bibinfo  {journal} {Phys. Rev. Research}\ }\textbf {\bibinfo
  {volume} {5}},\ \bibinfo {pages} {043044} (\bibinfo {year}
  {2023})}\BibitemShut {NoStop}\bibitem [{\citenamefont {Clark}(2026)}]{Clark2026}\BibitemOpen
  \bibfield  {author} {\bibinfo {author} {\bibfnamefont {D.~G.}\ \bibnamefont
  {Clark}},\ }\href {https://arxiv.org/abs/2610.12426} {\bibinfo {title}
  {Lyapunov spectrum of random neural networks}} (\bibinfo {year} {2026}),\
  \Eprint {https://arxiv.org/abs/2610.12426} {arXiv:2610.12426
  [cond-mat.dis-nn]} \BibitemShut {NoStop}\bibitem [{\citenamefont {Oseledets}(1968)}]{Oseledets1968}\BibitemOpen
  \bibfield  {author} {\bibinfo {author} {\bibfnamefont {V.~I.}\ \bibnamefont
  {Oseledets}},\ }\href@noop {} {\bibfield  {journal} {\bibinfo  {journal}
  {Trans. Moscow Math. Soc.}\ }\textbf {\bibinfo {volume} {19}},\ \bibinfo
  {pages} {197} (\bibinfo {year} {1968})}\BibitemShut {NoStop}\bibitem [{\citenamefont {Pesin}(1977)}]{Pesin1977}\BibitemOpen
  \bibfield  {author} {\bibinfo {author} {\bibfnamefont {Y.~B.}\ \bibnamefont
  {Pesin}},\ }\href@noop {} {\bibfield  {journal} {\bibinfo  {journal} {Russ.
  Math. Surv.}\ }\textbf {\bibinfo {volume} {32}},\ \bibinfo {pages} {55}
  (\bibinfo {year} {1977})}\BibitemShut {NoStop}\bibitem [{\citenamefont {Kaplan}\ and\ \citenamefont
  {Yorke}(1979)}]{KaplanYorke1979}\BibitemOpen
  \bibfield  {author} {\bibinfo {author} {\bibfnamefont {J.~L.}\ \bibnamefont
  {Kaplan}}\ and\ \bibinfo {author} {\bibfnamefont {J.~A.}\ \bibnamefont
  {Yorke}},\ }in\ \href {https://doi.org/10.1007/BFb0064319} {\emph {\bibinfo
  {booktitle} {Functional Differential Equations and Approximation of Fixed
  Points}}},\ \bibinfo {series} {Lecture Notes in Mathematics}, Vol.\ \bibinfo
  {volume} {730},\ \bibinfo {editor} {edited by\ \bibinfo {editor}
  {\bibfnamefont {H.-O.}\ \bibnamefont {Peitgen}}\ and\ \bibinfo {editor}
  {\bibfnamefont {H.-O.}\ \bibnamefont {Walther}}}\ (\bibinfo  {publisher}
  {Springer},\ \bibinfo {address} {Berlin},\ \bibinfo {year} {1979})\ pp.\
  \bibinfo {pages} {204--227}\BibitemShut {NoStop}\bibitem [{\citenamefont {Feinberg}\ and\ \citenamefont
  {Zee}(1997)}]{FeinbergZee1997}\BibitemOpen
  \bibfield  {author} {\bibinfo {author} {\bibfnamefont {J.}~\bibnamefont
  {Feinberg}}\ and\ \bibinfo {author} {\bibfnamefont {A.}~\bibnamefont {Zee}},\
  }\href@noop {} {\bibfield  {journal} {\bibinfo  {journal} {Nucl. Phys. B}\
  }\textbf {\bibinfo {volume} {504}},\ \bibinfo {pages} {579} (\bibinfo {year}
  {1997})},\ \Eprint {https://arxiv.org/abs/cond-mat/9703087}
  {arXiv:cond-mat/9703087} \BibitemShut {NoStop}\bibitem [{\citenamefont {Deninger}(2011)}]{Deninger2011}\BibitemOpen
  \bibfield  {author} {\bibinfo {author} {\bibfnamefont {C.}~\bibnamefont
  {Deninger}},\ }\href {https://doi.org/10.1515/crelle.2011.012} {\bibfield
  {journal} {\bibinfo  {journal} {J. Reine Angew. Math.}\ }\textbf {\bibinfo
  {volume} {651}},\ \bibinfo {pages} {165} (\bibinfo {year}
  {2011})}\BibitemShut {NoStop}\bibitem [{\citenamefont {Fuglede}\ and\ \citenamefont
  {Kadison}(1952)}]{FugledeKadison1952}\BibitemOpen
  \bibfield  {author} {\bibinfo {author} {\bibfnamefont {B.}~\bibnamefont
  {Fuglede}}\ and\ \bibinfo {author} {\bibfnamefont {R.~V.}\ \bibnamefont
  {Kadison}},\ }\href@noop {} {\bibfield  {journal} {\bibinfo  {journal} {Ann.
  Math.}\ }\textbf {\bibinfo {volume} {55}},\ \bibinfo {pages} {520} (\bibinfo
  {year} {1952})}\BibitemShut {NoStop}\bibitem [{\citenamefont {Helton}\ \emph {et~al.}(2007)\citenamefont {Helton},
  \citenamefont {Rashidi~Far},\ and\ \citenamefont {Speicher}}]{Helton2007}\BibitemOpen
  \bibfield  {author} {\bibinfo {author} {\bibfnamefont {J.~W.}\ \bibnamefont
  {Helton}}, \bibinfo {author} {\bibfnamefont {R.}~\bibnamefont
  {Rashidi~Far}},\ and\ \bibinfo {author} {\bibfnamefont {R.}~\bibnamefont
  {Speicher}},\ }\href {https://doi.org/10.1093/imrn/rnm086} {\bibfield
  {journal} {\bibinfo  {journal} {Int. Math. Res. Not.}\ }\textbf {\bibinfo
  {volume} {2007}},\ \bibinfo {pages} {rnm086} (\bibinfo {year} {2007})},\
  \Eprint {https://arxiv.org/abs/math/0703510} {arXiv:math/0703510}
  \BibitemShut {NoStop}\bibitem [{\citenamefont {Benettin}\ \emph {et~al.}(1980)\citenamefont
  {Benettin}, \citenamefont {Galgani}, \citenamefont {Giorgilli},\ and\
  \citenamefont {Strelcyn}}]{Benettin1980}\BibitemOpen
  \bibfield  {author} {\bibinfo {author} {\bibfnamefont {G.}~\bibnamefont
  {Benettin}}, \bibinfo {author} {\bibfnamefont {L.}~\bibnamefont {Galgani}},
  \bibinfo {author} {\bibfnamefont {A.}~\bibnamefont {Giorgilli}},\ and\
  \bibinfo {author} {\bibfnamefont {J.-M.}\ \bibnamefont {Strelcyn}},\ }\href
  {https://doi.org/10.1007/BF02128237} {\bibfield  {journal} {\bibinfo
  {journal} {Meccanica}\ }\textbf {\bibinfo {volume} {15}},\ \bibinfo {pages}
  {21} (\bibinfo {year} {1980})}\BibitemShut {NoStop}\end{thebibliography}
\end{document}